%% file: nvSetBWTE.tex
\documentclass{acmsiggraph}                     % final
\usepackage{multirow}
\usepackage{amsmath}
\usepackage{amssymb}
\usepackage{amsfonts}
\usepackage{eucal}
\usepackage[latin1]{inputenc}
\usepackage[normalem]{ulem}
\usepackage{tabulary}
\usepackage{graphicx}


\ifreviewelse {
    \newcommand{\TODO}[1]{{\textbf{TODO:} \textit{#1}}} % Enable to show todo tags
}{%
    \newcommand{\TODO}[1]{}                        % enable to hide todo tags
}

\usepackage{graphicx}
\usepackage{color}
\usepackage{wrapfig}
\usepackage{ifthen}
\usepackage[]{algorithm2e}

\usepackage{parskip}

\usepackage[labelfont=bf,textfont=it]{caption}

\title{A massively parallel algorithm for constructing the BWT of large string sets}
\author{Jacopo Pantaleoni\thanks{e-mail: jpantaleoni@nvidia.com}}

\begin{document}

\newcommand{\picresdir}{final}
%\newcommand{\picresdir}{proxies}

%% The ``\maketitle'' command must be the first command after the
%% ``\begin{document}'' command. It prepares and prints the title block.

\maketitle

%% Abstract section.

\abstract

We present a new scalable, lightweight algorithm to incrementally construct the BWT and FM-index of large string sets such as those produced by Next Generation Sequencing.
The algorithm is designed for massive parallelism and can effectively exploit the combination of low capacity high-bandwidth memory and slower external system memory typical of GPU accelerated systems. Particularly, for a string set of $n$ characters from an alphabet with $\sigma$ symbols, it uses a constant amount of high-bandwidth memory and at most $3n \log(\sigma)$ bits of system memory. Given that deep memory hierarchies are becoming a pervasive trait of high performance computing architectures, we believe this to be a relevant
feature.
The implementation can handle reads of arbitrary length and is up to 2 and respectively 6.5 times faster than state-of-the-art for short and long genomic reads.

\section{Introduction}

Recently, BWT and FM-index construction of very large string sets has become an important building block for bioinformatics applications such as \emph{de novo} assembly \cite{Simpson:2011} and compression of large genomic databases \cite{Cox:2012}.
In this context, a few novel lightweight algorithms have been developed to work specifically with very large collections of relatively short DNA strings, substantially outperforming previously known general purpose algorithms \cite{Bauer:2011,Li:2014}.
However, these algorithms are either serial or offer very limited parallelism.
Liu et al \shortcite{Liu:2014} has recently provided a new massively parallel algorithm that exploits the excellent sorting speed of modern GPUs for blockwise suffix sorting \cite{Karkkainen:2007} but the algorithm's speed is strongly limited by the speed at which suffixes can be gathered from the CPU's memory subsystem, which is several times slower than high bandwidth GPU memory. Moreover, Liu et al's \shortcite{Liu:2014} algorithm is not incremental.
This work provides a novel algorithm, {\bf set-bwte}, that drastically reduces the impact of the external
memory speed bottleneck while allowing to incrementally add new strings to pre-existing indices.
The algorithm can be seen as extending and adapting the {\bf bwte} algorithm by Ferragina et al \shortcite{Ferragina:2012} to string sets and modern parallel architectures with deep memory hierarchies.

\section{Overview}

Let $\Sigma = \{c_1, ..., c_\sigma\}$ be an ordered alphabet, with $c_1 < c_2 < \dots < c_\sigma$.
For a string $T = T[0 \dots n-1]$, we denote its $i$-th symbol with $T[i]$, and its $i$-th suffix with $T^i = T[i \dots n-1]$.

The \emph{suffix array} of $T$ is an array $SA[0 \dots n-1]$ such that $SA[i]$ is the index of the $i$-th smallest suffix of $T$, i.e. $\{ T^{SA[0]} \leq \dots \leq T^{SA[n-1]}\}$.
The \emph{Burrows-Wheeler Transform}, or \emph{BWT} of T is a string B defined as:
\begin{equation}
B[i] = T[ (SA[i] - 1) \mod n ]
\end{equation}
The BWT of a string set $(S_i)_{0 \leq i < m }$ is defined as the BWT of the string
$T = S_0 \$_0 \dots S_{m-1} \$_{m-1} $, where we define the special terminator symbols $(\$_i)_{0 \leq i < m }$ such that $\$_0 < \dots < \$_{m-1} < c_1$.

As typical in the treatment of the FM-index \cite{Ferragina:2005}, 
we define the following string \emph{ranking} operation:
\begin{equation}
rank(c,k,B) = |\{i < k : B[i] = c\}| 
\end{equation}
counting the occurrences of a character $c$ in the prefix $B[0 \dots k-1]$.

Similarly to the {\bf bwte} algorithm,
%developed by Ferragina et al \shortcite{Ferragina:2012}
{\bf set-bwte} partitions the input
set of strings in K blocks ${0 = j_0 < \dots < j_K = m}$, such that each block $S_{j_k} = S[j_k, j_{k+1})$ contains roughly the same
amount of suffixes $M$, and adds each block in turn to the partial BWT of the previously added blocks, $B_{ext}$.
This is done by first computing the SA of the new block to sort the suffixes relative to each other, then ranking the new, sorted suffixes
relative to $B_{ext}$, and finally inserting the corresponding BWT symbols in sorted order.
Pseudo-code is given in Algorithm 1.

\begin{algorithm}
\For{each block $S_{j_k}$} {
  $n          := j_{k+1} - j_k$\;
  $n_{suf} := \sum_{P \in S_{j_k}}{(1 + |P|)}-1$\;

  \emph{// build the suffix array of the block $S_{j_k}$ }\\
  $SA_{int}[0 \dots n_{suf}] :=$ {\bf ConstructSA}($S_{j_k}$)\;
  
  \emph{// extract the BWT symbols given the SA }\\
  $B_{int}[0 \dots n_{suf}] := B(S_{j_k}, SA_{int})$\;

  \emph{// rank the suffixes of $S_{j_k}$ in $B_{ext}$ }\\
  $g[0 \dots n_{suf}] :=$ {\bf ComputeRanks}($S_{j_k}, B_{ext}$)\;
  
  \emph{// reorder $g$ by the suffix order }\\
  %$g_{sa}[0 \dots n_{suf}] := \{ g[SA_{int}[i]]  : i \in [0,n_{suf}]\}$
  $g_{sa}[0 \dots n_{suf}] := \{ g[SA_{int}[0]] \dots g[SA_{int}[n_{suf}]]\}$

  \emph{// insert the symbols of $B_{int}$ in $B_{ext}$ at $g_{sa}$ }\\
  $B_{ext} := $ {\bf Insert}($B_{int}, g_{sa}, B_{ext}$)\;
}
\caption{set-bwte}
\end{algorithm}

Notice that the algorithm differs from {\bf bwte} for some relevant aspects: the first is that, unlike the single-string algorithm which proceeded backwards
from the end of the string, in {\bf set-bwte} the strings are added in a forward loop, starting from the beginning of the string set (though reversal is also possible).
This is a crucial difference, allowing the algorithm to be used for adding new strings at any time. This is possible because suffixes have limited length and
don't propagate across block boundaries.
The second is that rather than computing the ranks of the \emph{old} suffixes into the new block, we do the opposite, ranking the new ones relative to the external
BWT: this allows to perform only $O(|B_{int}|)$ work at each step, as opposed to $O(|B_{ext}|)$. %More details about this step are given in section 4.

\section{Suffix Array Construction}

Construction of the suffix array of a block of strings can be done with any suffix sorting algorithm.
Our implementation uses a massively parallel GPU based MSD radix-sort algorithm that treats strings as long integer keys made of multiple 32-bit words,
and sieves unique keys at each iteration in a spirit similar to that of the method described by Larsson and Sadakane \shortcite{Larsson:2007}.
The algorithm runs entirely in GPU memory as the blocks are limited in size.
On a Tesla K40, we use blocks of about 250 million symbols.

\section{Computing Ranks}

Computing the ranks of the suffixes of a string $P$ with respect to the external BWT can be done
with an adaptation of Lemma 1 of Ferragina et al \shortcite{Ferragina:2012}.
\paragraph{Lemma 1}: Let $C[c]$ denote the number of symbols in $B_{ext}$ that are smaller
than c, and suppose that suffix $P^k$ is lexicographically larger than precisely $i$ suffixes in $B_{ext}$.
Then, $P^{k-1}$ is lexicographically larger than precisely $j = C[c] + rank(c,i,B_{ext})$ suffixes, where
$c = P[k-1].$

This gives us a simple recipe for a massively parallel implementation of ComputeRanks, in Algorithm 2.
Notice that as anticipated this is reversing the roles of the new and old suffixes ($B_{int}$ and $B_{ext}$)
in the single-string version of the lemma provided by Ferragina et al \shortcite{Ferragina:2012}.
%We hence perform only $O(|B_{int}|)$ work at each step, as opposed to $O(|B_{ext}|)$.

\begin{algorithm}
\KwData{a block of $M$ strings $(P_i)_{0 \leq i <  M}$; $B_{ext}$}
\KwResult{g[$0 \dots \sum{(|P_j|+1})$]}
offsets[ $0 \dots M-1$ ]  := {\bf prefixsum}( $|P_j|+1$ )\;
\ForAll{j in $[0,M-1]$} {
  $k := |P_j|$\;
  $i := n_{ext}$\;
  offset := offsets[$j$]\;
  $g$[offset + $j + k$] := $i$\;
  \While {$k > 0$}{
  	$k := k - 1$\;
	$c := P_j[k]$\;
	$i := C[c] + rank(c,i,B_{ext})$\;
  	$g$[offset + $j + k$] := $i$\;
  }
}
\caption{ComputeRanks}
\end{algorithm}

\section{Insertion}

The last step of Algorithm 1 involves inserting the symbols $B_{int}[i]$ at the positions $g_{sa}[i]$ in $B_{ext}$. Unlike Ferragina et al \shortcite{Ferragina:2012}, who kept $B_{ext}$ on disk and used serials scans to perform the insertions, we keep $B_{ext}$ in system memory (i.e. the most \emph{external} layer in the random access memory hierarchy of a GPU accelerated system), and employ a new data structure allowing highly parallel insertion.

\renewcommand{\arraystretch}{1.3}

\begin{table}[ht] 
	\centering
{
\begin{tabulary}{1.0\textwidth}{|c|l|c|r|r|}
\hline
Data & Algorithm & RAM & Time & Throughput \\
\hline
NA12878 & beetl-bcr & 1.8G & 11.2h & 3.1 Mbp/s \\
\hline
NA12878 & ropebwt-bcr & 39.3G & 3.3h & 10.5 Mbp/s \\
\hline
NA12878 & ropebwt2 & 34.0G & 5.0h & 6.9 Mbp/s \\
\hline
NA12878 & nvSetBWT & 63.8G & 4.1h & 8.4 Mbp/s \\
\hline
NA12878 & {\bf set-bwte} & 78.0G &  {\bf 1.7h} & {\bf 20.4} Mbp/s \\
\hline
Venter & ropebwt2 & 22.2G & 1.4h & 5.4 Mbp/s \\
\hline
Venter & {\bf set-bwte} & 28.6G & {\bf 780s} & {\bf 35.5} Mbp/s \\
\hline
\end{tabulary}
}
\caption{\label{Table:Benchmarks} Benchmarks. Results have been generated on the following hardware: CPU: 24-core Xeon E5-2597-v2 at 2.7Ghz, GPU: NVIDIA Tesla K40, RAM: 128GB. }\vspace*{-0.2cm}
\end{table}

The data structure is essentially a paged array, with pages containing, at any time, between $p/2$ and $p$ symbols. Together with the actual page storage (allocated from a pool) we keep an index of the pages $P$, and an offset vector $O$ specifying the global offset of the first symbol of each page.
The latter allows to efficiently locate all the pages containing all the insertion points with a vectorized binary search.
Once such pages are located, they can be assigned to different threads in a pool which take care
of inserting the new symbols at the proper place and eventually split them if overflowing.
This data structure can be thought of as a \emph{flattened} B+ tree, where the hierarchy has been removed and replaced by a flat index in order to allow more memory-efficient parallel searches.
In our experiments, this provided tenfold speedups on modern multi-core architectures.

In order to facilitate ranking, we incorporate symbol occurrence counters in this paged array. Specifically, we mantain a set of $\sigma$ 64-bit global counters for each page,
and a sampled set of relative 32-bit counters within each page. For DNA we use a spacing of 128 symbols (resulting in 1 bit per symbol).

\section{Complexity Analysis}

In the following, we assume a set of $m$ strings totalling $n$ characters; we further denote with $l$ the average blockwise LCP length, i.e. $l = avg_{j_k}(LCP(S_{j_k}))$; notice that $l$ can be much smaller than the global LCP $L$ if the input is divided in many blocks.
Our parallel suffix sorting algorithm has a worst case complexity of $O(m l)$, though typically the average runtime is much better due to our suffix filtering strategy (which prunes sorted suffixes at an exponential rate on random input).
Ranking has complexity $O(n)$ and insertion has an asymptotic complexity of $O(n \log(n))$.
Hence, the algorithm has a total complexity of $O( m l + n \log(n) )$.
This is asymptotically lower than the $O( m L + n^2/M )$ limit of Liu et al's approach \shortcite{Liu:2014}.

In practice, for genomic datasets the $ml$ dependence is rather weak, and the algorithm seems fairly insensitive to the average read length.
It has also to be noted that these limits are only asymptotic: for finite $n$ and large block sizes $M$ (i.e. when $K = n / M$ is relatively small), the random bulk insertions touch each page, so that the complexity of insertion is initially proportional to $n^2 / M$. \footnote{for fixed $n$ the quantity $n^2/M$ is in fact an improvement over $p n \log(n)$, if the page size $p$ is larger than $n/M$; this also applies to other paged data structures, such as rope and B+ trees.} The actual cost of ranking is also superlinear in $n$, due to the fact that the memory accesses in $B_{ext}$ become sparser and sparser as $n$ grows. These factors typically dominate the runtime.

The algorithm uses a constant amount of high speed memory, and at most $3n \log(\sigma)$ bits of system memory: $2n \log(\sigma)$ bits for the paged array,
and up to $n\log(\sigma)$ bits for the occurrence counters.

\section{Results and Discussion}

We implemented the algorithm and evaluated its performance on the previously published NA12878 dataset containing 1.2 billion x 101bp human reads \cite{Depristo:2011}, and the Venter dataset containing 32 million x 875bp Sanger reads.
In Table 1 we compared the results to those published by Li \shortcite{Li:2014} on similar hardware, including four other algorithms: Illumina's beetle-bcr,
implementing the original disk based BCR \cite{Bauer:2011}; ropebwt-bcr, implementing an in-memory version of BCR; ropebwt2, implementing the algorithm by Li \shortcite{Li:2014}; and NVBIO's nvSetBWT 0.9.8, implementing a custom version of the algorithm by Liu et al \shortcite{Liu:2014}. Of these, only ropebwt2 can efficiently handle longer reads.

For short reads {\bf set-bwte} is almost 2 times faster than the previously fastest non-incremental algorithm, {ropebwt-bcr}, and almost 3 times faster than the state-of-the-art incremental algorithm, {ropebwt2}.
For longer reads the gap is even larger, reaching a speedup of 6.5x.

In our implementation, we perform suffix array construction on the GPU and ranking and insertion on the CPU. The three steps plus I/O are performed as stages in a pipeline,
where separate CPU threads execute each stage while performing multiple buffering of the outputs (see Figure 1). Thus, I/O, GPU and CPU computations are carefully overlapped.
Table 2 reports the average throughputs of the individual stages.
For large sets of short reads, on the Xeon E5-2597-v2 system we ran on, ranking was the bottleneck, and overall CPU processing took almost
three times as long as the GPU portion.
In other words a system with the same GPU and a faster CPU providing higher memory bandwidth could theoretically run almost 3 times as fast.

\begin{table}[ht] 
	\centering
{
\begin{tabulary}{1.0\textwidth}{|c|c|c|}
\hline
Stage & Venter & NA12878 \\
\hline
Suffix Sorting & 43 Mbp/s & 77 Mbp/s \\
\hline
Ranking & 57 Mbp/s & 37 Mbp/s \\
\hline
Insertion & 98 Mbp/s & 44 Mbp/s \\
\hline
\end{tabulary}
}
\caption{\label{Table:Breakdown} Breakdown of the throughputs of the various stages of our pipeline for the NA12878 and Venter datasets.}\vspace*{-0.2cm}
\end{table}

We would now like to draw the reader's attention on the fundamental primitives employed by our and other external memory BWT construction algorithms.
From a high-level perspective, {\bf set-bwte} can be thought of as a special form of \emph{insertion sort}.
Specifically, at each iteration the algorithm \emph{scatters} partially sorted  \emph{symbols} into an external BWT to avoid having to explicitly mantain
a full sorting index, which would demand impractical amounts of memory.
BCR \cite{Bauer:2011} falls into the same category and could be parallelized in a similar manner (except it requires working with a core dataset of exactly $m$ suffixes,
whereas our algorithm allows for an arbitrary block size).
This can be thought of as the dual of the blockwise suffix sorting \cite{Karkkainen:2007} skeleton employed by Liu et al \shortcite{Liu:2014},
which \emph{gathers} sparse input \emph{suffixes} from system memory.
Besides granting a lower asymptotic bound, we empirically observed that the former can be performed more efficiently on contemporary shared memory multi-core architectures,
where all cores share the overall memory system's bandwidth.
It has to be pointed out, however, that the latter is more scalable to small clusters, where each node can keep a copy of the entire input string-set and perform gathers
independently - whereas scattered insertion in a shared database would necessarily involve inter-node communication. We believe the superior single-node performance
and the incremental capability of our algorithm to be interesting tradeoffs against the better inter-node scalability of the slower, non-incremental blockwise suffix sorting
algorithms.

\begin{figure}
\begin{center}
 \includegraphics[width=75.0mm]{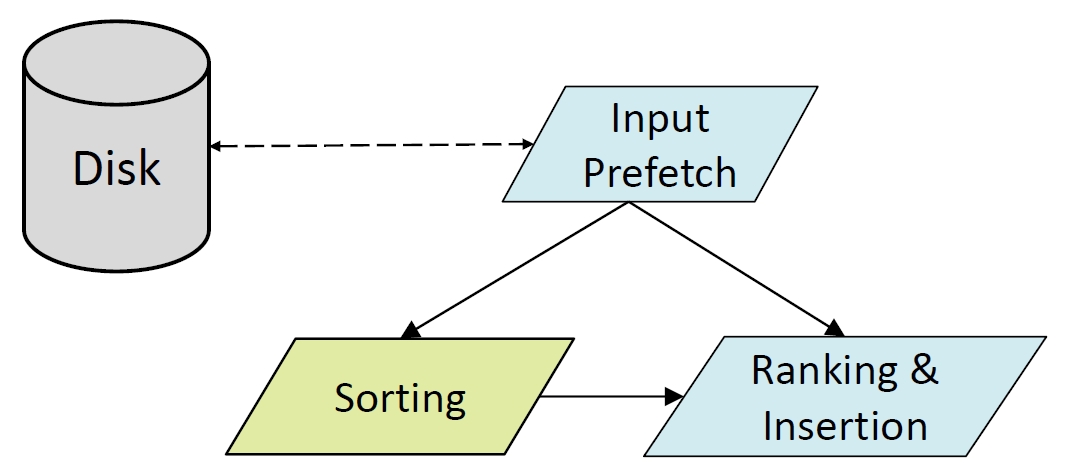}
 \caption{Our parallel pipeline visualized as a DAG. The blue boxes are stages performed by the CPU, the green box is performed on the GPU.}
\end{center}
\end{figure}

\section{Aknowledgements}

We thank Jonathan Cohen for insightful discussions and Heng Li for providing the benchmark datasets and initial feedback.

\bibliographystyle{acmsiggraph}
%\nocite{*}
\bibliography{main}
\end{document}